\documentclass[aps,prl,twocolumn,superscriptaddress,nofootinbib,floatfix]{revtex4-2}

\usepackage{amsmath,amssymb,bm}
\usepackage{physics}
\usepackage{hyperref}
\usepackage{mathtools}
\usepackage{xcolor}
\usepackage{ulem}

\hypersetup{
    colorlinks=true,
    linkcolor=blue,
    citecolor=blue,
    urlcolor=blue
}

\renewcommand{\vec}[1]{{\boldsymbol #1}}
\newcommand{\jj}{\vec j}
\newcommand{\EE}{\vec E}
\newcommand{\kk}{\vec k}
\newcommand{\zz}{\hat{\vec z}}

\begin{document}
\title{Anomalous Hall Effect Driven by Chiral Superconductivity}


\author{Alex Levchenko}
\affiliation{Department of Physics, University of Wisconsin-Madison, Madison, Wisconsin 53706, USA}

\author{Leonid Levitov}
\affiliation{Department of Physics, Massachusetts Institute of Technology, Cambridge MA02139, USA }

\date{\today}

\begin{abstract}
Direct dc-current signatures of unconventional superconductivity remain scarce. Existing probes of unconventional pairing are typically indirect, relying on phase-diagram anomalies, responses to external fields, or optical measurements. Here we propose a zero-field Hall drag effect as a direct transport signature of chiral superconductivity. The effect arises from Coulomb drag between quasiparticles in a chiral superconductor and those in an adjacent time-reversal-symmetric normal layer. We develop a minimal hydrodynamic theory that includes both quasiparticle normal current and condensate supercurrent in the superconducting layer. 
In an open-circuit superconducting layer, the condensate generates a counterflowing supercurrent that cancels the net layer current, while a finite quasiparticle current remains and mediates the transverse drag response. This results in anomalous Hall voltage signal appearing abruptly when $T$ is lowered below $T_c$, of the sign reflecting the sign of the superconducting order parameter phase winding.
\end{abstract}
\maketitle

Recent years have been marked by growing interest in novel forms of exotic superconductivity~\cite{Uchoa2007,BlackSchaffer2007,Honerkamp2008,Pathak2010,
Kiesel2012,Uchoa2013,Nandkishore2012}, an interest that accelerated after the discovery of superconductivity in twisted bilayer graphene~\cite{Cao2018MAGSC,
Chen2019TLG,
Lu2019MAG,
Stepanov2020MAG,
Yankowitz2019TBG,
Zhou2022BBG},
followed by correlated states that break time-reversal symmetry in rhombohedral graphene~\cite{Choi2025SCQAH,Guo2025ThickRMG,Han2023Multiferro,
Kumar2025DualSurface,
Yang2025SOCrmg,Zhou2021RTGSC,Zhou2022BBG}.
Most recently, direct signatures of chiral (time-reversal-broken) superconductivity have been reported
in rhombohedral graphene tetralayer and pentalayer~\cite{han2025chiral,choi2025qah}.
These developments 
prompt questions about new physical effects that can be used for probing such phases \cite{geier2026chiral,sau2024vortex,matsyshyn2024sBCD,daido2024hall,
messica2024halldrag,fu2025chern}.

Broken time-reversal symmetry (TRS) in metals is conventionally diagnosed through a Hall response, which may occur even in zero applied magnetic field when the electronic state has an intrinsic anomalous Hall conductivity. It is therefore natural to ask whether an analogous transport diagnostic can be formulated for chiral superconductors, whose order parameter also breaks TRS. The difficulty is that in a superconductor the condensate short-circuits the quasiparticle current in the dc limit, thereby obscuring both the ordinary Hall effect and any intrinsic anomalous Hall response of thermally excited Bogoliubov quasiparticles.

\begin{figure}[t]
    \centering
    \includegraphics[width=0.99\columnwidth]{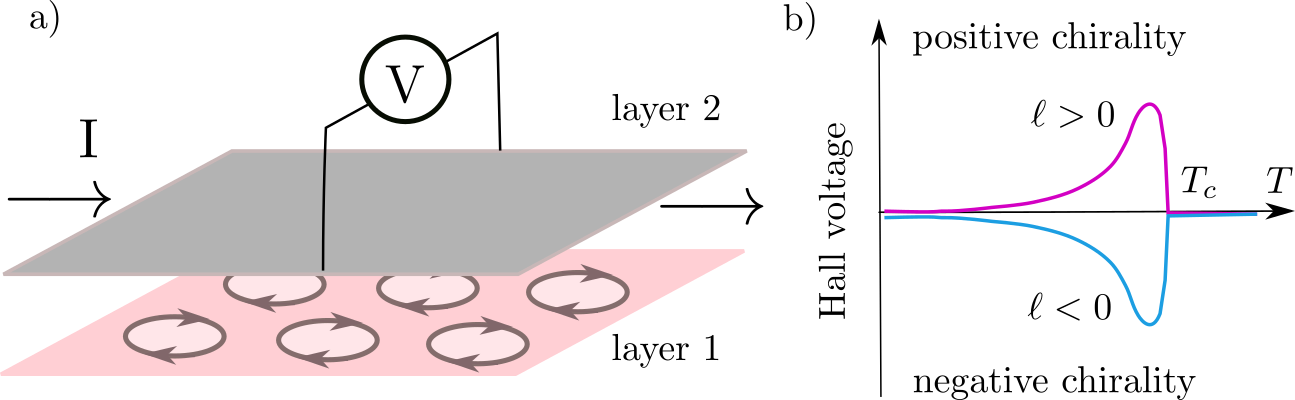}
    \caption{a) Zero-field Hall drag geometry: a chiral superconducting layer (1) and a normal layer (2) separated by a dielectric spacer. A longitudinal current is driven through the normal layer, while transverse probes measure the induced Hall voltage. The zero-field Hall response arises from interlayer Coulomb scattering involving the chiral quasiparticle response of the superconducting layer. b) Temperature dependence of Hall voltage, which is zero above $T_c$, grows rapidly below $T_c$ and drops exponentially at lower $T$ (see Eq.\eqref{eq:sigma_H_Tdependence}).
    }
    \vspace{-4mm}
    \label{fig}
\end{figure}

This limitation can be overcome by using Coulomb drag between a normal-metal layer and a nearby chiral-superconducting layer, as pictured in Fig.~\ref{fig}. A current applied in the normal layer drives its quasiparticles out of equilibrium, and interlayer Coulomb scattering transfers momentum to thermally excited quasiparticles in the superconducting layer at finite temperature $T<T_c$. This produces a quasiparticle current in the chiral superconductor even when the net electrical current in that layer is constrained to vanish by a compensating supercurrent.

Because the superconducting state is chiral, the Bogoliubov--de Gennes quasiparticle bands can carry Berry curvature, which generates a transverse anomalous velocity and hence a Hall component of the quasiparticle current. By reciprocal drag, this transverse quasiparticle current in the superconducting layer acts back on the normal layer and produces a drag-induced current or electric field at a nonzero Hall angle relative to the originally applied current. The resulting Hall voltage is measured in the normal layer, not directly in the superconductor, and is therefore not short-circuited by the superconducting condensate; see Fig.~\ref{fig}. Its onset below $T_c$, together with its reversal under reversal of chirality, would provide a `contactless' transport diagnostic of time-reversal symmetry breaking in the superconducting state.

The induced Hall resistivity is proportional to the square of the interlayer drag coefficient times the chiral quasiparticle Hall conductivity of the superconducting layer, $\rho_{xy}^{\rm drag}\sim \rho_D^2\sigma^{\rm qp}_{xy}$. The effect requires sufficiently strong drag coupling between the normal metal and the superconductor, a condition that is favorable in atomically thin and closely spaced conducting layers such as graphene-based heterostructures. Apart from this materials requirement, the mechanism is quite general: it relies only on finite-temperature superconducting quasiparticles, interlayer momentum transfer, and broken TRS in the superconducting quasiparticle spectrum.

Our microscopic 
picture of chiral quasiparticle transport driven by the drag electric field $\EE_1^{(D)}(\omega)$ 
is an extension of nonchiral quasiparticle kinetics 
derived by Pethick and Smith (PS) \cite{PethickSmith1979}. It can be summed up as:
\begin{equation}
    \left(\partial_t+\tau_1^{-1}\right)
    \left(
        \jj_{1n}-\sigma^{\rm qp}_{xy}\,\zz\times\EE_1^{(D)}
    \right)
    =
    D_1\EE_1^{(D)},
    \label{eq:chiral_drude_time}
\end{equation}
Here, \(D_1=n_{n1}e^2/m\) is the normal Drude weight of layer 1, \(\tau_1\) is the quasiparticle momentum relaxation time, and \(\sigma^{\rm qp}_{xy}\) is the anomalous Hall conductivity of the chiral superconducting quasiparticles (see Eqs.~\eqref{eq:sigma_chi}, \eqref{eq:PS565} and \eqref{eq:sigmaL_belyaev}). The Hall term is placed inside the brackets to indicate that the relaxation operator acts on the drift current only; the transverse term is an anomalous-velocity contribution to the quasiparticle current, not an additional longitudinal driving force.

The frequency dependence is introduced to highlight the dynamics of the drag response; throughout we assume that the driving frequency \(\omega\) is small compared with the superconducting gap scale, \(\hbar\omega\ll \Delta\), so that quasiparticle transport is quasistatic on the scale of pair breaking.
In frequency space,
\begin{equation}
    \jj_{1n}(\omega)
    =
    \sigma_{1L}(\omega)\EE_1^{(D)}(\omega)
    +
    \sigma^{\rm qp}_{xy}(\omega)\,\zz\times\EE_1^{(D)}(\omega),
    \label{eq:sigma_tensor_sc}
\end{equation}
where $\sigma_{1L}$ and $\sigma^{\rm qp}_{xy}$ are the quasiparticle longitudinal and transverse conductivities discussed below (see Eqs.~\eqref{eq:sigma_chi}, \eqref{eq:sigmaL_belyaev}).

Note that the drag field $\EE^{(D)}$ is not an externally applied electromagnetic field in the superconducting layer. Rather, it phenomenologically represents the interlayer Coulomb momentum-transfer force acting on drag-active quasiparticles. Direct transfer of this incoherent momentum to the uniform condensate is suppressed: as a phase-coherent collective mode, the condensate momentum can change only through a coherent phase twist, not through relaxation-time scattering. The condensate responds indirectly by generating the supercurrent required to satisfy particle conservation, $\jj_n+\jj_s=0$.

Next, we summarize the microscopic origin of the transverse quasiparticle conductivity term $\sigma^{\rm qp}_{xy}$ in a two-dimensional chiral superconductor. 
A minimal single-band Bogoliubov--de Gennes (BdG) Hamiltonian with a chiral gap $\Delta_{\kk}=|\Delta_{\kk}|e^{i\ell\phi_{\kk}}$, where $\ell$ is phase winding,  is
\begin{equation}
    H_{\rm BdG}(\kk)=\mathbf d(\kk)\cdot\bm\tau,
    \quad
    \mathbf d(\kk)=
    \left(
        \Re\Delta_{\kk},
        -\Im\Delta_{\kk},
        \xi_{\kk}
    \right).
\end{equation}
The Berry curvature of this band is
\begin{equation}
    \Omega_{\lambda,z}(\kk)
    =
    \lambda\frac{1}{2}
    \hat{\mathbf d}\cdot
    \left(
        \partial_{k_x}\hat{\mathbf d}
        \times
        \partial_{k_y}\hat{\mathbf d}
    \right),
    \label{eq:berry_general}
\end{equation}
where $\lambda=\pm$ is BdG band index, \(\hat{\mathbf d}=\mathbf d/|\mathbf d|\). For an isotropic chiral state this reduces, up to convention-dependent signs, to
\begin{equation}
    \Omega_{+,z}(k)
    =
    \frac{\ell}{2k}\frac{d}{dk}
    \left(\frac{\xi_k}{E_k}\right).
\end{equation}
The curvature is concentrated near the Fermi surface, where electron-hole mixing is strongest.

The semiclassical quasiparticle equations of motion are 
\begin{align}
    \dot{\mathbf r}
    =
    \frac{1}{\hbar}\nabla_{\kk}E_{\kk}
    -\dot{\kk}\times\bm\Omega_{\kk},
    \quad
    \hbar\dot{\kk}
    =
    e q_{\kk}\EE
    ,
\end{align}
\cite{wang2021berry,liao2023chiral,liao2024spin}, 
where $q_{\kk}$ is Bogoliubov effective quasiparticle charge
$
    q_{\kk}=u_{\kk}^2-v_{\kk}^2 
    .
$
Consequently,
$
    \dot{\mathbf r}
    =
    \mathbf v_{\kk}
    +
    \frac{e q_{\kk}}{\hbar}\Omega_z(\kk)\,\zz\times\EE
$, where the second term is the anomalous velocity. The effective charge value $q_{\bm{k}}$ follows from BdG algebra. Because the quasiparticle is only partially electron-like, the electric field acts on it with the reduced charge $q_{\bm{k}}$, and the Berry-curvature anomalous velocity is then built on the same $q_{\bm{k}}$. 
Since both the force and the transported charge are proportional to  $q_{\bm{k}}$, the quasiparticle anomalous Hall conductivity has the form
\begin{equation}
    \sigma^{\rm qp}_{xy}
    =
    \frac{e^2}{\hbar}
    \sum_\lambda
    \int\frac{d^2k}{(2\pi)^2}
    f_{\lambda\kk}\,
    q_{\lambda\kk}^{2}\,
    \Omega_{\lambda,z}(\kk),
    \label{eq:sigma_chi}
\end{equation}
where \(f_{\lambda\kk}\) is the local quasiparticle distribution. This expression should be understood as the intrinsic Berry-curvature contribution to the quasiparticle Hall response. 
The role of other contributions to Hall conductivity chirality-induced impurity scattering---side-jump, skew scattering and related extrinsic Hall/Kerr mechanisms
\cite{Goryo2008,LutchynNagornykhYakovenko2009,LiAndreevSpivak2015,
KonigLevchenko2017,LiuChenHuang2023}---will be discussed later.

This yields a characteristic $T$ dependence $\sigma^{\rm qp}_{xy}(T)$: 
an abrupt  rise 
from zero at $T>T_c$ 
 to $ \sigma^{\rm qp}_{xy}(T)\sim \frac{\Delta(T)}{T_c} \, \frac{\ell}6\frac{e^2}{h}$ at $T\lesssim T_c$, followed by exponential drop at lower $T$. This behavior is illustrated in Fig.\,\ref{fig}b).
A compact interpolation that captures this $T$ dependence is
\begin{equation}
\label{eq:sigma_H_Tdependence}
\sigma^{\rm qp}_{xy}(T)=\frac1{e^{\Delta(T)/T}+1}\,\frac{\Delta(T)}{T}\,\frac{\ell}3\frac{e^2}{h}
\end{equation}
These equations apply to a chiral superconducting state under the following assumptions: 
Breaking of TRS does not by itself invalidate the 
PS longitudinal normal-current equation.  Rather, chirality enlarges the hydrodynamic response by allowing a transverse, time-reversal-odd Hall sector.  The longitudinal part of Eq.~\eqref{eq:chiral_drude_time} remains applicable provided the same hydrodynamic assumptions hold and the chiral order parameter can be treated as a fixed broken-symmetry background.

Theoretical description of quasiparticle dynamics in 
a chiral superconductor, must account for various effects originating from the gap phase winding around the Fermi surface and the Bogoliubov quasiparticle bands carrying Berry curvature.  In addition, gap anisotropy, nodes, impurity-induced interbranch scattering, skew scattering, side-jump terms, and collective dynamics of chirality domains can all modify the simple scalar relaxation-time approximation. The minimal extension adopted here, Eq.~\eqref{eq:chiral_drude_time}, keeps the PS longitudinal Drude sector and adds an intrinsic anomalous Hall current generated by the quasiparticle Berry curvature.

In this sense, the chiral generalization, Eq.\eqref{eq:chiral_drude_time}, should be read as longitudinal 
PS dynamics supplemented by a TRS-odd anomalous Hall response.  The resulting equations are intended for a homogeneous chiral domain, long wavelengths, and frequencies low enough for hydrodynamics but high enough, or with enough relaxation, that the passive superconducting boundary conditions can be imposed consistently.

Next, we 
express the chiral quasiparticle response to a nonchiral normal layer through 
the interlayer drag coupling and 
`integrate out' the superconducting-layer variables to obtain the induced Hall field. Let layer 1 be the chiral superconducting layer and layer 2 the nonchiral normal layer. The normal layer has scalar conductivity \(\sigma_2\), or resistivity \(\rho_2=1/\sigma_2\). The scalar 
drag coupling is taken to be
\begin{equation}
    \EE_1^D=\rho_D\jj_2,
    \qquad
    \EE_2^D=\rho_D\jj_{1n}.
    \label{eq:drag_law_qp}
\end{equation}
The second relation is written in terms of the quasiparticle current in the superconducting layer, not the net current. This is essential. If one instead couples drag to the net current \(\jj_{1,\rm tot}\), an open-circuit superconducting layer with \(\jj_{1,\rm tot}=0\) would trivially produce no back-action drag field.

To unravel the role of supercurrents in the drag problem, it is instructive to begin with the nonchiral case, where drag is purely longitudinal. Current in layer 1, mediated by interlayer quasiparticle scattering, can transfer momentum to quasiparticles in the superconducting layer more easily than to the condensate. Indeed, since the typical kinetic energy of drag-related scattering processes is small (of order $k_B T$) and quasiparticles in a superconductor are gapped, excitation out of the condensate into the normal subsystem is suppressed. In this case, the drag field accelerates only quasiparticles, not the condensate. The response of the condensate in this case can be understood by analogy with the fountain effect in superfluids: a phase gradient builds up, generating a supercurrent that is equal and opposite to the quasiparticle current. This counterflow regime is characterized by zero net current and by a phase difference in the superconducting layer induced by drag from the normal layer.

Next, we restore chirality and add details to this picture. The quasiparticle current induced in layer 1 by the drag field from layer 2 is
\begin{equation}
    \jj_{1n}
    =
    \rho_D\sigma_{1L}\jj_2
    +
    \rho_D\sigma^{\rm qp}_{xy}\,\zz\times\jj_2 .
    \label{eq:j1n_from_j2}
\end{equation}
The two terms are respectively the longitudinal drag-induced quasiparticle current and the transverse chiral Hall current.
The reciprocal drag field back in the normal layer is
\begin{equation}
    \EE_2^D
    =
    \rho_D\jj_{1n}
    =
    \rho_D^2\sigma_{1L}\jj_2
    +
    \rho_D^2\sigma^{\rm qp}_{xy}\,\zz\times\jj_2 .
    \label{eq:E2D_backaction}
\end{equation}
Adding the ordinary electric field of the normal layer gives
\begin{equation}
    \EE_2
    =
    \rho_2\jj_2+
        \rho_D^2\sigma_{1L}\jj_2
    +
    \rho_D^2\sigma^{\rm qp}_{xy}\zz\times\jj_2.
    \label{eq:E2_effective}
\end{equation}
For a current \(\jj_2=j_2\hat{\vec x}\), the transverse component of Eq.~\eqref{eq:E2_effective} gives the induced Hall resistivity in the otherwise nonchiral layer, 
\begin{equation}
    \rho_{xy}^{(2)}
    =
    \frac{E_{2y}}{j_{2x}}
    =
    \rho_D^2\sigma^{\rm qp}_{xy} .
    \label{eq:main_result}
\end{equation}
The corresponding Hall angle is
\begin{equation}
    \tan\theta_H^{(2)}
    =
    \frac{E_{2y}}{E_{2x}}
    =
    \frac{\rho_D^2\sigma^{\rm qp}_{xy}}
    {\rho_2+
        \rho_D^2\sigma_{1L}}.
\end{equation}
The effect is odd under reversal of the chirality, since
$    \sigma^{\rm qp}_{xy}(+\ell)=-\sigma^{\rm qp}_{xy}(-\ell)$.
A nonzero chiral drag and its sign can therefore serve as a direct signature of phase winding in the paired state. 

We can now impose the passive-superconductor boundary condition and show that it does not remove the quasiparticle current that mediates drag. 
The appropriate boundary condition is 
that the net electrical current in layer 1 must vanish:
$    \jj_{1,\rm tot}=\jj_{1n}+\jj_{1s}=0$.
This condition does not require the quasiparticle current itself to vanish. Instead, the condensate develops a counterflowing supercurrent,
\begin{equation}
    \jj_{1s}=-\jj_{1n}.
    \label{eq:counterflow}
\end{equation}
In London form,
\begin{equation}
    \jj_{1s}
    =
    \frac{n_{s1}e}{m}
    \left(
        \hbar\nabla\varphi_1-\frac{2e}{c}\mathbf A_1
    \right).
\end{equation}
For negligible vector potential, this fixes the phase gradient:
\begin{equation}
    \frac{n_{s1}e\hbar}{m}\nabla\varphi_1
    =
    -\jj_{1n}
    =
    -\rho_D\sigma_{1L}\jj_2
    -\rho_D\sigma^{\rm qp}_{xy}\,\zz\times\jj_2 .
    \label{eq:phase_gradient}
\end{equation}
Thus the superconducting layer can satisfy the closed-circuit or open-circuit net-current boundary condition while retaining a finite quasiparticle current. Since the interlayer drag force acts on the quasiparticle/normal component, the back-action field remains $ \EE_2^D=\rho_D\jj_{1n}$ as in Eq. \eqref{eq:drag_law_qp}, rather than \(\rho_D\jj_{1,\rm tot}\). Therefore, the anomalous Hall drag signal, Eq.~\eqref{eq:main_result}, survives in the presence of condensate counterflow.

We then test the robustness of this counterflow picture by adding a minimal normal-condensate conversion process. 
The open-circuit analysis above assumes that a finite quasiparticle current can coexist with a compensating supercurrent, so that the total current in the superconducting layer vanishes.  It is useful to ask whether this picture is stable against Belyaev-like relaxation processes, by which the condensate and the quasiparticle fluid exchange particles.  In superconductors the more standard terminology is quasiparticle recombination, pair breaking, branch-imbalance relaxation, and gap relaxation.  The phenomenology is closely related to the Rothwarf-Taylor rate equations for quasiparticle recombination and pair-breaking phonons, microscopic electron--phonon calculations of quasiparticle and phonon lifetimes by Kaplan, et al., and the kinetic treatments of branch imbalance and gap relaxation by Schmid-Schon and Pethick-Smith~\cite{RothwarfTaylor1967,Kaplan1976,SchmidSchon1975Kinetic,PethickSmith1979}.  The name ``Belyaev-like'' is used here only as an analogy to condensate-normal-cloud exchange in finite-temperature Bose condensates, where Zaremba-Nikuni-Griffin-type kinetic theories contain explicit condensate-noncondensate collision integrals and source terms in the condensate continuity equation~\cite{ZarembaGriffinNikuni1998,ZarembaNikuniGriffin1999}.  For the present superconducting drag problem, we model these microscopic processes by a conversion term proportional to the relative chemical potential, while treating its coefficient as a phenomenological relaxation rate.  Such processes do not violate total charge conservation.  Rather, they relax the relative population and chemical-potential imbalance between the normal and superfluid components.  A minimal two-fluid model is
\begin{align}
    \partial_t n_n+\nabla\cdot \jj_n
    &=
    -\Gamma_B \mu_- ,
    \\
    \partial_t n_s+\nabla\cdot \jj_s
    &=
    +\Gamma_B \mu_- ,
    \label{eq:belyaev_continuity}
\end{align}
where
\begin{equation}
    n=n_n+n_s,
    \qquad
    \mu_- = \mu_n-\mu_s .
\end{equation}
Adding the two equations gives
$
    \partial_t n+
    \nabla\cdot(\jj_n+\jj_s)=0
$.
Thus conversion transfers density between the normal and condensate sectors, but preserves the total density and total charge.  In the charge-neutral open-circuit limit,
\begin{equation}
    \delta n\simeq 0,
    \qquad
    \jj_n+\jj_s\simeq 0,
\end{equation}
such that the net current condition is automatically satisfied.  The conversion term drives \(\mu_n-\mu_s\to 0\), analogous to chemical-potential equilibration in finite-temperature two-fluid kinetic theory, on a time scale
$
    \tau_B \sim \frac{\chi_-}{\Gamma_B},
$
where \(\chi_-\) is the relative compressibility.  It suppresses long-lived branch imbalance, but it does not by itself force \(\jj_n=0\).

The simplest way to include this relaxation in the normal-current dynamics is to let it contribute to the damping of relative normal-superfluid motion.  For the chiral superconducting layer,
\begin{equation}
    \jj_{1n}
    =
    \sigma_{1L}^{B}(\omega)\EE_1^D
    +
    \sigma_{xy}^{B}(\omega)\zz\times \EE_1^D,
    \label{eq:belyaev_current_response}
\end{equation}
with
    $\EE_1^D=\rho_D\jj_2$.
A minimal longitudinal response is
\begin{equation}
    \sigma_{1L}^{B}(\omega)
    =
    \frac{D_{1n}}
    {\tau_{\rm imp}^{-1}+\tau_B^{-1}-i\omega},
    \label{eq:sigmaL_belyaev}
\end{equation}
$D_{1n}=n_{1n}e^2/m$.
The chiral Hall coefficient \(\sigma^{\rm qp}_{xy}(\omega)\) denotes the corresponding Hall response after including the same conversion physics.  If the Hall current is dominated by a nonequilibrium drift distortion of the quasiparticle distribution, then \(\tau_B^{-1}\) enters its effective relaxation rate.  If the Hall current is dominated by an intrinsic Berry-curvature anomalous velocity of thermally occupied Bogoliubov quasiparticles, conversion mainly renormalizes its magnitude rather than forcing it to vanish.

The condensate still adjusts its phase to enforce the open-circuit condition,
$\jj_{1s}=-\jj_{1n}$,
$\jj_{1,\rm tot}=0$, 
provided the superfluid stiffness is finite and phase adjustment is fast on the time scale of the drive.  The reciprocal drag field in the normal layer is then still controlled by the quasiparticle current,
$
    \EE_2^D=\rho_D\jj_{1n},
$
not by the vanishing total current.  Combining this with Eq.~\eqref{eq:belyaev_current_response} gives
\begin{equation}
    \EE_2^D
    =
    \rho_D^2
    \left[
        \sigma_{1L}^{B}(\omega)\jj_2
        +
        \sigma_{xy}^{B}(\omega)\zz\times\jj_2
    \right].
\end{equation}
Therefore the measured field in the normal layer becomes
\begin{equation}
    \EE_2
    =
    \rho_2\jj_2
    +
    \rho_D^2\sigma_{1L}^{B}(\omega)\jj_2
    +
    \rho_D^2\sigma_{xy}^{B}(\omega)\zz\times\jj_2.
    \label{eq:belyaev_drag_field}
\end{equation}
The induced Hall resistivity is consequently
\begin{equation}
    \rho_{xy}^{(2)}(\omega)
    =
    \rho_D^2\sigma_{xy}^{B}(\omega),
    \label{eq:belyaev_hall_result}
\end{equation}
while the longitudinal drag correction is
$   \rho_{xx}^{(2)}(\omega)
    =
    \rho_2+
    \rho_D^2\sigma_{1L}^{B}(\omega)$. 
The counterflow mechanism is therefore robust in the following sense: normal-condensate conversion renormalizes and damps the counterflow susceptibility, but it does not remove the Hall drag signal as long as the drag-active quasiparticle Hall conductivity remains finite.  The effect is weakened in the strong-conversion limit if \(\sigma^{\rm qp}_{xy}\) is entirely tied to a drift nonequilibrium distribution.  It survives as a renormalized response if the chiral layer retains an intrinsic Berry-curvature Hall conductivity of Bogoliubov quasiparticles.

We estimate the anomalous Hall drag value for a graphene double layer system (for details, see Appendix B). Taking a representative parameter regime ($E_F = 10\text{ meV}$, $\Delta_0 = 0.2\text{ meV}$, and $r_s = 4.5$) at finite temperature, with $T \approx 0.09\text{ meV}$ and $\Delta(T) \approx 0.14\text{ meV}$, we obtain $\rho_{xy}^D \sim 20\text{ m}\Omega$. This value is well within the detection limit of modern lock-in transport measurements.

Our analysis uses the standard Coulomb-drag framework, in which interlayer momentum transfer arises to second order in the interlayer interaction and is governed by nonlinear charge-current susceptibilities of the two layers~\cite{Pogrebinskii1977,Price1983,JauhoSmith1993,KamenevOreg1995,Rojo1999,NarozhnyLevchenko2016}.  In ordinary zero-field Fermi-liquid bilayers this response is longitudinal; Hall drag requires time-reversal symmetry breaking, for example by a magnetic field, a quantum Hall state, or a time-reversal-odd electronic structure.  Superconducting drag has also been studied, particularly by Kamenev and Oreg, who showed that superconducting fluctuations can strongly enhance drag near a transition~\cite{KamenevOreg1995}.  Our mechanism is distinct: the transverse drag response is generated not by an external field or ordinary fluctuations, but by the intrinsic chirality of the superconducting quasiparticle spectrum, producing a zero-field Hall signal in an adjacent time-reversal-invariant normal layer.

In contrast to an ordinary metal, the uniform electrical Hall conductivity of a chiral superconductor is a subtle object: in a clean Galilean-invariant single-band model the intrinsic dc charge Hall response is strongly constrained, while multiband structure, finite frequency, disorder, vertex corrections, and impurity-induced subgap bands can generate anomalous electromagnetic Hall effects~\cite{Goryo2008,YangWu2018,TaylorKallin2012,LiWangHuang2020}.
Yet, the predicted anomalous Hall effect remains robust in the presence of chirality-induced impurity scattering---side-jump, skew scattering and related extrinsic Hall/Kerr mechanisms
\cite{Goryo2008,LutchynNagornykhYakovenko2009,LiAndreevSpivak2015,
KonigLevchenko2017,LiuChenHuang2023}.

A closely related, and in some respects cleaner, setting is thermal Hall transport.  Bogoliubov quasiparticles carry heat without the same charge-conservation constraints, and anomalous zero-field thermal Hall transport can arise from branch-conversion scattering by the chiral order parameter and impurity potential~\cite{NgampruetikornSauls2020,NgampruetikornSaulsReview2024}.

Another 
interesting precedent is the anomalous Hall transport of electron bubbles and ions in chiral superfluid $^3$He-A~\cite{IkegamiScience2013,IkegamiJPSJ2015,IkegamiMobility2013}. The microscopic theory attributes the transverse force to skew scattering of quasiparticles from the chiral order parameter around an electron bubble~\cite{ShevtsovSaulsPRB2016,ShevtsovSaulsJLTP2017}, building on earlier calculations of ion mobility tensors and possible transverse terms in $^3$He-A~\cite{SalomaaPethickBaym1980,SalmelinPRL1989,SalmelinPRB1990}. This body of work provides a close conceptual analogue: a nominally non-superfluid probe object acquires a chirality-dependent transverse response through scattering from chiral quasiparticles. Similar to the observations of a chiral Hall effect in superfluid $^3$He \cite{IkegamiScience2013,IkegamiJPSJ2015,IkegamiMobility2013}, we expect chiral drag to appear abruptly when $T$ is lowered below $T_c$, with its sign reflecting the sign of the superconducting order parameter phase winding.

In summary, this work identifies a zero-field Hall drag response as a direct transport signature of chiral superconductivity. The Hall signal is not a direct dc response of the superconducting layer, but a drag-induced voltage in the adjacent normal layer arising from the chiral quasiparticle response of the superconductor. Open-circuit conditions do not suppress the effect: the condensate supercurrent cancels the net current, while the quasiparticle current that mediates drag remains finite. The resulting Hall voltage is odd under chirality reversal and turns on below $T_c$, providing a direct transport signature of time-reversal symmetry breaking.

This work greatly benefited from discussions with Maxim Dzero, Chris Pethick, James Sauls, Boris Spivak, Long Ju and Andrea Young. The work of A. L. was financially supported by National Science Foundation Grant No. DMR-2452658 and H. I. Romnes Faculty Fellowship provided by the University of Wisconsin-Madison Office of the Vice Chancellor for Research and Graduate Education with funding from the Wisconsin Alumni Research Foundation.


\appendix
\section{Appendix A: Quasiparticle kinetics}


\setcounter{equation}{0}


To add context to the discussion of the chiral problem in the main text, we briefly recall the Pethick-Smith analysis of the two-fluid normal-current equation from which the longitudinal Drude sector is obtained. 
In this analysis, the quasiparticle contribution to the normal current and its effect on supercurrent
must be distinguished \cite{PethickSmith1979,PethickSmith1980}.  This distinction is important for the drag problem because the condensate can provide a counterflowing supercurrent, whereas interlayer Coulomb drag primarily acts on the quasiparticle/normal component.  The broader nonequilibrium-superconductivity context includes branch-imbalance work by Tinkham and Clarke, the microscopic dirty-limit kinetic equations of Schmid and Sch\"on, phenomenological two-fluid collective-mode theory by Bray and Schmidt, and clean-limit and unconventional-gap extensions by Artemenko, Volkov, and Kobelkov~\cite{TinkhamClarke1972,SchmidSchon1975Kinetic,SchmidSchon1975Collective,BraySchmidt1975,ArtemenkoVolkov1975Collective,ArtemenkoVolkov1979Review,ArtemenkoKobelkov1997}.

As a starting point, we recall that a Bogoliubov quasiparticle carries the effective charge
\begin{equation}
    q_{\kk}=u_{\kk}^2-v_{\kk}^2=\frac{\xi_{\kk}}{E_{\kk}},
\end{equation}
where
\begin{equation}
    E_{\kk}=\sqrt{\xi_{\kk}^2+|\Delta_{\kk}|^2}.
\end{equation}
Consequently, the current of quasiparticle charge is not generally identical to the usual normal current.

The normal current $\jj_n$ appearing in the Pethick--Smith equation is the usual two-fluid normal current: it is the contribution of thermally excited Bogoliubov quasiparticles to the physical particle/electrical current.  The coefficient $\rho_n$ is the corresponding normal-fluid density or stiffness, defined by the two-fluid decomposition
\begin{equation}
    \jj=\jj_s+\jj_n,\qquad
    \jj_s=\rho_s\mathbf v_s,
    \qquad
    \rho_s+\rho_n=\rho .
\end{equation}
As a cautionary remark, $\rho_n$ is distinct from the density of Bogoliubov quasiparticles counted as excitations, nor it is the quasiparticle charge susceptibility.

Let $f_\kk$ be the Bogoliubov quasiparticle distribution and let $f^{\rm le}_\kk$ be the local-equilibrium distribution set by the instantaneous condensate variables, such as $\mu_s$, $\Delta$, and $\mathbf v_s$.  The nonequilibrium correction is
\begin{equation}
    \delta f_\kk\equiv f_\kk-f^{\rm le}_\kk .
\end{equation}
In this notation, the quasiparticle charge density, quasiparticle-charge current, and usual normal current have the schematic forms
\begin{align}
    Q_n
    &=
    \sum_\kk q_\kk\,\delta f_\kk,
    \\
    \jj_n^Q
    &=
    \sum_\kk q_\kk\mathbf v_\kk\,\delta f_\kk,
    \\
    \jj_n
    &=
    \sum_\kk \frac{\hbar\kk}{m}\,\delta f_\kk .
\end{align}
Spin factors and normalization of the momentum sums are suppressed.  The distinction is that $\jj_n^Q$ weights quasiparticles by their effective charge $q_\kk$, whereas $\jj_n$ weights them by the mechanical current carried by the excitation.  Thus $Q_n$ and $\jj_n^Q$ describe charge imbalance of the quasiparticle component, while $\jj_n$ is the normal current entering ordinary electrodynamics and the Drude-like equation.

Two common 
perturbed distributions are useful to keep separate.  A quasiparticle charge-imbalance mode has
\begin{equation}
    \delta f_\kk
    \simeq
    -q_\kk\delta\mu\,\partial_E f_0(E_\kk),
\end{equation}
corresponding to a quasiparticle chemical potential shifted relative to the condensate.  A current-carrying drift distortion has
\begin{equation}
    \delta f_\kk
    \simeq
    -\hbar\kk\cdot\mathbf v_n\,\partial_E f_0(E_\kk),
\end{equation}
which produces the usual normal current $\jj_n$.

The derivation of the Drude/Ohm equation follows from the kinetic equation by taking the normal-current moment. Technically, this represents  
a hydrodynamic reduction of the kinetic theory of nonequilibrium superconductors: one starts from a gauge-invariant quasiparticle kinetic equation, or equivalently from the Eilenberger--Gor'kov--Eliashberg formulation in the dirty limit, separates the distribution into a local-equilibrium part set by the condensate variables and a small nonequilibrium correction, and takes the moment of the kinetic equation weighted by the mechanical current of a Bogoliubov excitation.  The resulting normal-current equation is coupled to the condensate through the normal and superfluid charge susceptibilities.  In the long-wavelength Carlson--Goldman/Artemenko--Volkov regime, Coulomb screening enforces approximate charge neutrality, so the pressure and chemical-potential-gradient terms combine into a term proportional to $\nabla(Q_n+Q_s)$ and drop out.  Impurity momentum relaxation is then represented by a relaxation-time collision integral, giving the longitudinal Drude response of the normal component.  This logic is closely related to the Schmid--Sch\"on microscopic dirty-limit treatment, the Bray--Schmidt phenomenological two-fluid model, the clean-limit Artemenko--Volkov collective-mode theory, and later quasiclassical treatments of unconventional superconductors such as the Artemenko--Kobelkov $d$-wave analysis~\cite{SchmidSchon1975Kinetic,SchmidSchon1975Collective,BraySchmidt1975,ArtemenkoVolkov1975Collective,ArtemenkoKobelkov1997}.

The normal-current equation is obtained by multiplying the quasiparticle Boltzmann equation by the 
current carried by a quasiparticle and summing over states.  Before the final cancellation, the structure is
\begin{align}
    \partial_t\jj_n
    &+
    \frac{v_F^2}{3}\nabla Q_n
    +
    \left(
        \frac{\rho_n}{m}
        -
        \frac{v_F^2}{3}\chi_n^0
    \right)\nabla\mu_s
    \nonumber \\
    &+
    \rho_n\frac{e}{m}\nabla\Phi
    =
    \left(\partial_t\jj_n\right)_{\rm coll} .
    \label{eq:jn_before_cancel}
\end{align}
Here $\mu_s$ is the condensate chemical potential, $\Phi$ is the electrostatic potential, and $\chi_n^0$ is the bare normal-component charge susceptibility.  Near $T_c$, to the relevant order in $\Delta/k_BT_c$,
\begin{equation}
    \frac{\rho_n}{m}
    -
    \frac{v_F^2}{3}\chi_n^0
    \simeq
    \frac{v_F^2}{3}\chi_s^0,
\end{equation}
while the superfluid charge susceptibility gives
\begin{equation}
    Q_s=\chi_s^0\mu_s .
\end{equation}
Therefore the two gradient terms combine as
\begin{equation}
    \frac{v_F^2}{3}\nabla Q_n
    +
    \frac{v_F^2}{3}\chi_s^0\nabla\mu_s
    =
    \frac{v_F^2}{3}\nabla(Q_n+Q_s).
\end{equation}
In the long-wavelength collective-mode regime, Coulomb screening suppresses total charge fluctuations,
\begin{equation}
    Q_n+Q_s\simeq 0,
\end{equation}
so this pressure/compressibility-gradient contribution drops out.  With impurity momentum relaxation treated by
\begin{equation}
    \left(\partial_t\jj_n\right)_{\rm coll}
    =
    -\frac{\jj_n}{\tau_{\rm imp}},
\end{equation}
and $\EE=-\nabla\Phi$, Eq.~\eqref{eq:jn_before_cancel} reduces to the Pethick--Smith Drude/Ohm relation
\begin{equation}  
\partial_t\jj_n-\rho_n(e/m)\EE=-\jj_n/\tau_{\rm imp}.
\end{equation}  
For the physical electrical current $\vec J_n=e\jj_n$, this becomes
\begin{equation}
\label{eq:PS565}
    \left(\partial_t+\tau_{\rm imp}^{-1}\right)\vec J_n
    =
    \frac{\rho_n e^2}{m}\vec E .
\end{equation}

\medskip
\noindent The resulting equation has a specific hydrodynamic domain of validity. 

Eq.~\eqref{eq:PS565} is a hydrodynamic, long-wavelength, near-$T_c$ reduction.  Its use assumes:
\begin{enumerate}
    \item well-defined Bogoliubov quasiparticles described by a Boltzmann equation;
    \item variations slow compared with the gap-adjustment scale $\hbar/\Delta$ and over distances long compared with the coherence length;
    \item a controlled expansion near $T_c$, so that the susceptibility identities used above are valid to the required order in $\Delta/k_BT_c$;
    \item long-wavelength charge neutrality, $Q_n+Q_s\simeq 0$, enforced by Coulomb screening;
    \item impurity-dominated momentum relaxation represented by a single relaxation time $\tau_{\rm imp}$;
    \item no additional slow order-parameter variables beyond the condensate phase and amplitude included in the two-fluid description.
\end{enumerate}
Fermi-liquid corrections can be included, but in the charge-neutral collective mode the effective-field terms proportional to the total charge fluctuation cancel to this order.

\section{Appendix B: Magnitude estimate}

For two clean two-dimensional Fermi liquids, the temperature dependence of Coulomb drag at $T\ll E_F$ is parametrically given by
\begin{equation}
\rho_D \simeq \frac{h}{e^2} A_{12}(k_F d) \left(\frac{T}{E_F}\right)^2,
\label{eq:normal_drag}
\end{equation}
up to numerical factors of order unity. Here, $A_{12}$ is determined by the matrix elements of the interlayer interaction. The precise form of $A_{12}$ as a function of the interlayer spacing $d$ and Fermi momentum $k_F$ depends on the strength of the coupling. In the strong-coupling regime ($r_s\gtrsim 1$), for closely spaced layers ($k_Fd<1$) it becomes essentially independent of $k_Fd$ and scales as the square of the characteristic interlayer coupling constant, $A_{12}\propto r_s^2$. 


This describes, in particular, graphene double-layer systems 
at particle density of the order of $n \sim 2 \times 10^{12}\text{ cm}^{-2}$ and  graphene layer spacing $d \lesssim 3.5\text{ nm}$ or smaller. For these parameter values, the product $k_Fd$ satisfies the inequality $k_F d < 1$. These parameters are close to those describing graphene layers separated by a thin hBN spacer. 

In a two-dimensional electron system, the dimensionless parameter $r_s$ measures the ratio of the inter-particle Coulomb repulsion energy ($E_C$) to the kinetic energy ($E_K$, which is typically on the order of the Fermi energy $E_F$). It is defined as $r_s = E_C/E_K = e^2 / (\kappa a E_F)$, where $a = 1/\sqrt{\pi n}$ is the average inter-particle distance for a 2D density $n$, and $\kappa$ is the effective background dielectric constant. This estimate yields $r_s \sim 4.5$. We recall that in a standard parabolic band system, $r_s$ can be written using the effective mass $m^*$ as $r_s \propto m^* / (\kappa \sqrt{n})$. However, in the context of rhombohedral graphene layers, the displacement field flattens the electronic states at the perimeter of the M-shaped valence band; the local effective mass $m^*$ increases significantly, thus enhancing the effective $r_s$. Because an analytical expression for $\rho_D$ is usually derived via perturbation theory in the interaction, Eq. \eqref{eq:normal_drag} cannot be straightforwardly applied when $r_s \gg 1$. For this reason, the estimate we provide below for the anomalous Hall drag resistivity serves as a lower-bound estimate.

Using the interpolation formula for $\sigma^{\text{qp}}_{xy}$, we can present the induced anomalous Hall drag resistivity in the form:
\begin{equation}
\rho_{xy}^D \sim \frac{h}{e^2} r_s^4 \left(\frac{T}{E_F}\right)^4 \frac{\Delta}{T} \frac{1}{e^{\Delta/T}+1}
\end{equation}
Optimizing this expression with respect to $\Delta/T$ yields $x = \Delta/T \approx 1.55$. Taking a specific parameter regime ($E_F = 10\text{ meV}$, $\Delta_0 = 0.2\text{ meV}$, and $r_s = 4.5$) at a finite temperature where $T \approx 0.09\text{ meV}$ and $\Delta(T) \approx 0.14\text{ meV}$, we arrive at $\rho_{xy}^D \sim 20\text{ m}\Omega$. This magnitude is well within the detectable resolution of modern lock-in transport setups.


\begin{thebibliography}{99}

\bibitem{Uchoa2007}
B. Uchoa and A.~H. Castro Neto,
``Superconducting states of pure and doped graphene,''
Phys.\ Rev.\ Lett.\ \textbf{98}, 146801 (2007).

\bibitem{BlackSchaffer2007}
A.~M. Black-Schaffer and S. Doniach,
``Resonating valence bonds and mean-field d-wave superconductivity in graphite,''
Phys.\ Rev.\ B \textbf{75}, 134512 (2007).

\bibitem{Honerkamp2008}
C. Honerkamp,
``Density waves and Cooper pairing on the honeycomb lattice,''
Phys.\ Rev.\ Lett.\ \textbf{100}, 146404 (2008).

\bibitem{Pathak2010}
S. Pathak, V. B.~Shenoy, and G.~Baskaran,
``Possible high-temperature superconducting state with a d+id pairing symmetry in doped graphene,''
Phys.\ Rev.\ B \textbf{81}, 085431 (2010).

\bibitem{Nandkishore2012}
R. Nandkishore, L.~S. Levitov, and A.~V. Chubukov,
``Chiral superconductivity from repulsive interactions in doped graphene,''
Nat.\ Phys.\ \textbf{8}, 158--163 (2012).

\bibitem{Kiesel2012}
M. L.~Kiesel, C. Platt, W. Hanke, D. A.~Abanin, and R. Thomale,
``Competing many-body instabilities and unconventional superconductivity in graphene,''
Phys.\ Rev.\ B \textbf{86}, 020507 (2012).

\bibitem{Uchoa2013}
B.~Uchoa and Y.~Barlas,
``Superconducting states in pseudo-Landau-levels of strained graphene,''
Phys.\ Rev.\ Lett.\ \textbf{111}, 046604 (2013).

\bibitem{Cao2018MAGSC}
Y.~Cao, V.~Fatemi, S.~Fang, K.~Watanabe, T.~Taniguchi, E.~Kaxiras, and P.~Jarillo-Herrero,
``Unconventional superconductivity in magic-angle graphene superlattices,''
\textit{Nature} \textbf{556}, 43--50 (2018),

\bibitem{Yankowitz2019TBG}
M.~Yankowitz, S.~Chen, H.~Polshyn, Y.~Zhang, K.~Watanabe, T.~Taniguchi, D.~Graf, A.~F.~Young, and C.~R.~Dean,
``Tuning superconductivity in twisted bilayer graphene,''
\textit{Science} \textbf{363}, 1059--1064 (2019),

\bibitem{Lu2019MAG}
X.~Lu, P.~Stepanov, W.~Yang, M.~Xie, M.~A.~Aamir, I.~Das, C.~Urgell, K.~Watanabe, T.~Taniguchi, G.~Zhang, A.~Bachtold, A.~H.~MacDonald, and D.~K.~Efetov,
``Superconductors, orbital magnets and correlated states in magic-angle bilayer graphene,''
\textit{Nature} \textbf{574}, 653--657 (2019),

\bibitem{Chen2019TLG}
G.~Chen, A.~L.~Sharpe, P.~Gallagher, I.~T.~Rosen, E.~J.~Fox, L.~Jiang, B.~Lyu, H.~Li, K.~Watanabe, T.~Taniguchi, J.~Jung, Z.~Shi, D.~Goldhaber-Gordon, Y.~Zhang, and F.~Wang,
``Signatures of tunable superconductivity in a trilayer graphene moiré superlattice,''
\textit{Nature} \textbf{572}, 215--219 (2019),

\bibitem{Stepanov2020MAG}
P.~Stepanov, I.~Das, X.~Lu, A.~Fahimniya, K.~Watanabe, T.~Taniguchi, F.~H.~L.~Koppens, J.~Lischner, L.~Levitov, and D.~K.~Efetov,
``Untying the insulating and superconducting orders in magic-angle graphene,''
\textit{Nature} \textbf{583}, 375--378 (2020),





\bibitem{Zhou2022BBG}
H.~Zhou, L.~Holleis, Y.~Saito, L.~Cohen, W.~Huynh, C.~L.~Patterson, F.~Yang, T.~Taniguchi, K.~Watanabe, and A.~F.~Young,
``Isospin magnetism and spin-polarized superconductivity in Bernal bilayer graphene,''
\textit{Science} \textbf{375}, 774--778 (2022),




\bibitem{Choi2025SCQAH}
Y.~Choi \textit{et al.},
``Superconductivity and quantized anomalous Hall effect in rhombohedral graphene,''
Nature \textbf{639}, 342--347 (2025).

\bibitem{Zhou2021RTGSC}
H.~Zhou \textit{et al.},
``Superconductivity in rhombohedral trilayer graphene,''
Nature \textbf{598}, 434--438 (2021).

\bibitem{Guo2025ThickRMG}
Y.~Guo \textit{et al.},
``Flat band surface state superconductivity in thick rhombohedral graphene,''
arXiv:2511.17423 
(2025).


\bibitem{Kumar2025DualSurface}
M.~Kumar \textit{et al.},
``Superconductivity from dual-surface carriers in rhombohedral graphene,''
arXiv:2507.18598 [cond-mat.mes-hall] (2025).


\bibitem{Yang2025SOCrmg}
J.~Yang \textit{et al.},
``Impact of spin-orbit coupling on superconductivity in rhombohedral graphene,''
Nat.\ Mater.\ \textbf{24}, 1058-1065 (2025).

\bibitem{Han2023Multiferro}
T.~Han \textit{et al.},
``Orbital multiferroicity in pentalayer rhombohedral graphene,''
Nature \textbf{623}, 41-47 (2023).



\bibitem{han2025chiral} T. Han et al., “Signatures of chiral superconductivity in rhombohedral graphene,” Nature 643, 654-661 (2025).


\bibitem{choi2025qah} Y. Choi et al., “Superconductivity and quantized anomalous Hall effect in rhombohedral graphene,” Nature 639, 342–347 (2025).

\bibitem{geier2026chiral} M. Geier, M. Davydova, and L. Fu, “Chiral and topological superconductivity in isospin polarized multilayer graphene,” Nature Communications 17, 232 (2026; published online 2025).

\bibitem{sau2024vortex} J. D. Sau and S. Wang, “Theory of anomalous Hall effect from screened vortex charge in a phase disordered superconductor,” arXiv:2411.08969 [cond-mat.supr-con] (2024).

\bibitem{matsyshyn2024sBCD} O. Matsyshyn, G. Vignale, and J. C. W. Song, “Superconducting Berry Curvature Dipole,” arXiv:2410.21363 [cond-mat.supr-con] (2024).

\bibitem{daido2024hall} A. Daido and Y. Yanase, “Rectification and nonlinear Hall effect by fluctuating finite-momentum Cooper pairs,” Phys. Rev. Research 6, L022009 (2024).


\bibitem{messica2024halldrag} Y. Messica and D. B. Gutman, “Hall Coulomb drag induced by electron-electron skew scattering,” Phys. Rev. B 110, 115424 (2024).

\bibitem{fu2025chern} Yu Fu, Yu Huang, and Q. L. He, “Non-reciprocal Coulomb drag between Chern insulators,” Nature Communications 16, 3058 (2025).

\bibitem{wang2021berry}
Z. Wang, L. Dong, C. Xiao, and Q. Niu,
``Berry Curvature Effects on Quasiparticle Dynamics in Superconductors,''
Phys. Rev. Lett. \textbf{126}, 187001 (2021).

\bibitem{liao2023chiral}
Y. Liao and Y.-T. Hsu,
``Unveiling Quasiparticle Berry Curvature Effects in the Spectroscopic Properties of a Chiral p-wave Superconductor,''
arXiv:2311.02165 (2023).

\bibitem{liao2024spin}
Z.-C. Liao, C. Xiao, Z. Wang, and Q. Niu,
``Berry Curvature Induced Spin Nernst and Thermal Edelstein Effects in Proximity Superconductors,''
arXiv:2412.08451 (2024).

\bibitem{Bardeen1958}
J.~Bardeen,
``Two-Fluid Model of Superconductivity,''
Phys. Rev. Lett. \textbf{1}, 399-400 (1958).

\bibitem{TinkhamClarke1972}
M.~Tinkham and J.~Clarke,
``Theory of Pair-Quasiparticle Potential Difference in Nonequilibrium Superconductors,''
Phys. Rev. Lett. \textbf{28}, 1366-1369 (1972).

\bibitem{SchmidSchon1975Kinetic}
A.~Schmid and G.~Sch\"on,
``Linearized Kinetic Equations and Relaxation Processes of a Superconductor Near $T_c$,''
J. Low Temp. Phys. \textbf{20}, 207-227 (1975).

\bibitem{SchmidSchon1975Collective}
A.~Schmid and G.~Sch\"on,
``Collective Oscillations in a Dirty Superconductor,''
Phys. Rev. Lett. \textbf{34}, 941-943 (1975).

\bibitem{BraySchmidt1975}
A.~J.~Bray and H.~Schmidt,
``Collective Modes in Charged Superconductors Near $T_c$,''
Solid State Communications \textbf{17}, 1175-1178 (1975).

\bibitem{ArtemenkoVolkov1975Collective}
S.~N.~Artemenko and A.~F.~Volkov,
``Collective Excitations with a Sound Spectrum in Superconductors,''
Soviet Physics JETP \textbf{42}, 896-898 (1975)
[Zh. Eksp. Teor. Fiz. \textbf{69}, 1764--1767 (1975)].

\bibitem{ArtemenkoVolkov1975Stationary}
S.~N.~Artemenko and A.~F.~Volkov,
``Stationary Electric Field in Superconductors with Nonzero Energy Gap,''
Physics Letters A \textbf{55}, 113--114 (1975).

\bibitem{ArtemenkoVolkov1979Review}
S.~N.~Artemenko and A.~F.~Volkov,
``Electric Fields and Collective Oscillations in Superconductors,''
Soviet Physics Uspekhi \textbf{22}, 295--310 (1979).

\bibitem{PethickSmith1979}
C.~J.~Pethick and H.~Smith,
``Relaxation and Collective Motion in Superconductors: A Two-Fluid Description,''
Annals of Physics \textbf{119}, 133--169 (1979).

\bibitem{PethickSmith1980}
C.~J.~Pethick and H.~Smith,
``Charge Imbalance in Nonequilibrium Superconductors,''
Journal of Physics C: Solid State Physics \textbf{13}, 6313--6338 (1980).

\bibitem{PethickSmith1979PRL}
C.~J.~Pethick and H.~Smith,
``Generation of Charge Imbalance in a Superconductor by a Temperature Gradient,''
Physical Review Letters \textbf{43}, 640--643 (1979).

\bibitem{NicolskyFrotaPessoa1982}
R.~Nicolsky and S.~Frota-Pess\^oa,
``Voltage in a Charge-Imbalance Experiment, Taking into Account the Approximate Charge Neutrality in the Superconductor,''
J. Low Temp. Phys. \textbf{46}, 107--113 (1982).

\bibitem{ArtemenkoKobelkov1997}
S.~N.~Artemenko and A.~G.~Kobelkov,
``Linear Response and Collective Oscillations in Superconductors with $d$-Wave Pairing,''
Phys. Rev. B \textbf{55}, 9094--9104 (1997).


\bibitem{RothwarfTaylor1967}
A.~Rothwarf and B.~N.~Taylor,
``Measurement of Recombination Lifetimes in Superconductors,''
Phys. Rev. Lett. \textbf{19}, 27-30 (1967).

\bibitem{Kaplan1976}
S.~B.~Kaplan, C.~C.~Chi, D.~N.~Langenberg, J.-J.~Chang, S.~Jafarey, and D.~J.~Scalapino,
``Quasiparticle and Phonon Lifetimes in Superconductors,''
Phys. Rev. B \textbf{14}, 4854-4873 (1976).



\bibitem{ZarembaGriffinNikuni1998}
E.~Zaremba, A.~Griffin, and T.~Nikuni,
``Two-Fluid Hydrodynamics for a Trapped Weakly Interacting Bose Gas,''
Phys. Rev. A \textbf{57}, 4695-4707 (1998).

\bibitem{ZarembaNikuniGriffin1999}
E.~Zaremba, T.~Nikuni, and A.~Griffin,
``Dynamics of Trapped Bose Gases at Finite Temperatures,''
J. Low Temp. Phys. \textbf{116}, 277-345 (1999).


\bibitem{Goryo2008}
J.~Goryo,
``Impurity-induced polar Kerr effect in a chiral $p$-wave superconductor,''
Phys. Rev. B \textbf{78}, 060501(R) (2008).

\bibitem{LutchynNagornykhYakovenko2009}
R.~M. Lutchyn, P.~Nagornykh, and V.~M. Yakovenko,
``Frequency and temperature dependence of the anomalous Hall conductivity in a chiral $p_x+i p_y$ superconductor with impurities,''
Phys. Rev. B \textbf{80}, 104508 (2009).

\bibitem{LiAndreevSpivak2015}
S.~Li, A.~V. Andreev, and B.~Z. Spivak,
``Anomalous transport phenomena in $p_x+i p_y$ superconductors,''
Phys. Rev. B \textbf{92}, 100506(R) (2015).

\bibitem{KonigLevchenko2017}
E.~J. K\"onig and A.~Levchenko,
``Kerr effect from diffractive skew scattering in chiral $p_x \pm i p_y$ superconductors,''
Phys. Rev. Lett. \textbf{118}, 027001 (2017).

\bibitem{LiuChenHuang2023}
H.-T. Liu, W.~Chen, and W.~Huang,
``Impact of random impurities on the anomalous Hall effect in chiral superconductors,''
Phys. Rev. B \textbf{107}, 224517 (2023).

\bibitem{YangWu2018}
F.~Yang and M.~W.~Wu,
``Gauge-Invariant Microscopic Kinetic Theory of Superconductivity in Response to Electromagnetic Fields,''
Phys. Rev. B \textbf{98}, 094507 (2018).

\bibitem{Pogrebinskii1977}
M. B. Pogrebinskii,
``Mutual drag of carriers in a semiconductor--insulator--semiconductor system,''
Sov. Phys. Semicond. \textbf{11}, 372 (1977).

\bibitem{Price1983}
P. J. Price,
``Hot electrons in semiconductor heterolayers,''
Physica B+C \textbf{117--118}, 750 (1983).

\bibitem{JauhoSmith1993}
A.-P. Jauho and H. Smith,
``Coulomb drag between parallel two-dimensional electron systems,''
Phys. Rev. B \textbf{47}, 4420 (1993).

\bibitem{KamenevOreg1995}
A. Kamenev and Y. Oreg,
``Coulomb drag in normal metals and superconductors: Diagrammatic approach,''
Phys. Rev. B \textbf{52}, 7516 (1995).

\bibitem{Rojo1999}
A. G. Rojo,
``Electron-drag effects in coupled electron systems,''
J. Phys.: Condens. Matter \textbf{11}, R31 (1999).

\bibitem{NarozhnyLevchenko2016}
B. N. Narozhny and A. Levchenko,
``Coulomb drag,''
Rev. Mod. Phys. \textbf{88}, 025003 (2016).




\bibitem{IkegamiScience2013}
H. Ikegami, Y. Tsutsumi, and K. Kono, ``Chiral symmetry breaking in superfluid $^3$He-A,'' Science \textbf{341}, 59--62 (2013).

\bibitem{IkegamiJPSJ2015}
H. Ikegami, Y. Tsutsumi, and K. Kono, ``Observation of Intrinsic Magnus Force and Direct Detection of Chirality in Superfluid $^3$He-A,''J. Phys. Soc. Jpn. \textbf{84}, 044602 (2015).

\bibitem{IkegamiMobility2013}
H. Ikegami, S. B. Chung, and K. Kono, ``Mobility of Ions Trapped Below a Free Surface of Superfluid $^3$He,'' J. Phys. Soc. Jpn. \textbf{82}, 124607 (2013).

\bibitem{ShevtsovSaulsPRB2016}
O. Shevtsov and J. A. Sauls, ``Electron bubbles and Weyl fermions in chiral superfluid $^3$He-A,'' Phys. Rev. B \textbf{94}, 064511 (2016).

\bibitem{ShevtsovSaulsJLTP2017}
O. Shevtsov and J. A. Sauls, ``Electron Bubbles in Superfluid $^3$He-A: Exploring the Quasiparticle--Ion Interaction,'' J. Low Temp. Phys. \textbf{187}, 340--353 (2017).

\bibitem{SalomaaPethickBaym1980}
M. Salomaa, C. J. Pethick, and G. Baym, ``Mobility tensor of negative ions in superfluid $^3$He-A,'' J. Low Temp. Phys. \textbf{40}, 297--356 (1980).

\bibitem{SalmelinPRL1989}
R. H. Salmelin, M. M. Salomaa, and V. P. Mineev, Phys. Rev. Lett. \textbf{63}, 868 (1989).

\bibitem{SalmelinPRB1990}
R. H. Salmelin, M. M. Salomaa, and V. P. Mineev, Phys. Rev. B \textbf{41}, 4142 (1990).

\bibitem{TaylorKallin2012}
E. Taylor and C. Kallin,
``Intrinsic Hall effect in a multiband chiral superconductor in the absence of an external magnetic field,''
Phys. Rev. Lett. \textbf{108}, 157001 (2012).

\bibitem{LiWangHuang2020}
Y. Li, Z. Wang, and C.-K. Huang,
``Anomalous Hall effect in single-band chiral superconductors from impurity superlattices,''
Phys. Rev. Research \textbf{2}, 042027(R) (2020).

\bibitem{NgampruetikornSauls2020}
V. Ngampruetikorn and J. A. Sauls,
``Impurity-induced anomalous thermal Hall effect in chiral superconductors,''
Phys. Rev. Lett. \textbf{124}, 157002 (2020).

\bibitem{NgampruetikornSaulsReview2024}
V. Ngampruetikorn and J. A. Sauls,
``Anomalous Hall effects in chiral superconductors,''
Front. Phys. \textbf{12}, 1384275 (2024).
\end{thebibliography}
\end{document}